\documentclass[fleqn,usenatbib]{mnras}

\usepackage{newtxtext,newtxmath}
\usepackage[T1]{fontenc}
\usepackage{ae,aecompl}
\usepackage{xcolor} 

\usepackage[T1]{fontenc}

\DeclareRobustCommand{\VAN}[3]{#2}
\let\VANthebibliography\thebibliography
\def\thebibliography{\DeclareRobustCommand{\VAN}[3]{##3}\VANthebibliography}

\usepackage{graphicx}	
\usepackage{amsmath}	
\usepackage{tabularx}
\usepackage{hyperref}
\usepackage{comment}
\usepackage{dblfloatfix}
\usepackage{tablefootnote}

\title[Galactic Carbon Stars]{The frequency by mass of Galactic carbon stars inferred from Gaia measurements of star cluster members}

\author[T. Pal \& G. Worthey]{
Tathagata Pal$^{1}$\thanks{E-mail: tathagata.pal@wsu.edu},
G. Worthey$^{1}$
\\
$^{1}$Washington State University, Pullman, WA 99164, USA\\
}

\date{Accepted 2021 July 06. Received 2021 July 06; in original form 2020 December 11}

\pubyear{2021}

\begin{document}
\label{firstpage}
\pagerange{\pageref{firstpage}--\pageref{lastpage}}
\maketitle

\begin{abstract}
We investigate the frequency of occurrence of Galactic carbon stars as a function of progenitor mass using {\it Gaia} data. Small number statistics limit fidelity, but C~star frequency agrees with that observed in the Magellanic Clouds (MCs) down to $m \approx1.67$ M$_\odot$. At $m \approx 1.38$ M$_\odot$, the frequency rises by a factor of three even though the frequency appears to drop to zero for the MCs. In fact this is due to a lack of clusters at the key age range in the MCs. At $m \approx 1.24$ M$_\odot$ and below, no C stars are observed, corresponding to ages older than 4 Gyr. Within uncertainties, C~star frequency in M~31 is consistent with that of the Galaxy and the MCs. We find an ambiguous C-star candidate at $\sim$7 M$_\odot$.
\end{abstract}



\section{Introduction}


Late in the evolution of intermediate mass stars, carbon produced in the interior can mix to the surface and alter the chemical mixture. When the number density of carbon atoms dominates over oxygen, carbon stars (C stars) are formed whose red spectra are dominated by carbon-bearing molecules \citep{1986ApJ...303...10M}. We restrict our purview to these stars, known as classical carbon stars, the products of single-star evolution sampled late on the asymptotic giant branch (AGB) and generally among the N \citep{1928LicOB..13..123S} or C-N \citep{1993PASP..105..905K} spectral types, although extreme C stars can enshroud themselves with dust and disappear from the optical altogether \citep{1998ARA&A..36..369W,2005ARA&A..43..435H,2010JApA...31..177L}. 

C stars are very luminous and can carry a significant fraction of the integrated luminosity of a stellar population \citep{1998MNRAS.300..872M,2005MNRAS.362..799M} and are therefore important for integrated light studies of galaxies. Conclusions regarding integrated light contributions are based upon stellar evolutionary calculations, based in turn on carbon star counts in the Magellanic Clouds (MCs). Cluster ages \citep{1995A&A...298...87G} combined with star classifications and counts \citep{1990ApJ...352...96F} provide useful constraints on the number of carbon stars created as a function of age \citep{1996A&A...316L...1M,2007A&A...462..237G}. 

Counting carbon stars in M31, a closer analog to the Milky Way than the MCs, has been fraught with uncertainty due to image crowding and the search for a suitable diagnostic to distinguish between M stars and C stars of the same color \citep{2013ApJ...774...83B,2003AJ....125.2473S,2005AJ....129..201D}. These problems appear to have been resolved by \cite{2019ApJ...879..109B}, at least in terms of C/M ratio \cite{2005A&A...434..657B}. A C/M number ratio, however, only predicts the rate of C~star production if the expected number of M~giants can be estimated. That requires exact knowledge of the underlying age and metallicity distribution because metal-rich populations produce M~giants on the first-ascent red giant branch at ever-lower luminosities and greater numbers \citep[e.g.,][]{1994ApJS...95..107W}. We do not yet have this knowledge for M~31 except for its low-metallicity outer portions \citep[e.g.,][]{2003ApJ...592L..17B}.

Theoretical approaches struggle to produce ab initio predictions due to uncertainties about mass loss and other prescriptions. For a while, models \citep{2008A&A...482..883M} predicted large numbers of C stars in metal-rich populations, but this has reversed, and models now predict that at some uncertain supersolar metallicity a superwind terminates evolution before sufficient C dredge-up can occur \citep{2009A&A...508.1343W,2013MNRAS.434..488M}, so that C star production may have a metallicity ceiling.

Analysis similar to that done in the MCs \citep[e.g.,][]{2019MNRAS.485.5666P,2020MNRAS.498.3283P} for Milky Way C stars is difficult because, although thousands of C stars are known \citep{2001BaltA..10....1A}, almost none have known ages or masses on an individual basis.  \cite{2006MNRAS.369..791F} argued from kinematics that C-stars that are also Mira variables have mass $1.8 \pm 0.2$ M$_\odot$, but this subset is probably not applicable to every C star on the AGB. No individual parallaxes were known until HIPPARCOS and {\it Gaia} \citep{1998A&A...338..209A}, but even if a distance, and therefore an absolute magnitude, is known, assigning an individual mass is usually impossible \citep{2020A&A...633A.135A}.

We wondered if this situation could be improved, and therefore investigated carbon stars in open clusters (OCs) and dissipated clusters using {\it Gaia} data. The {\it Gaia} data release 2 (DR2) provides us with a wealth of information on astrometry and photometry data for about 1.3 billion galactic stars \citep{2018A&A...616A...8A}. Precise colour and magnitude data determines the age of clusters via isochrone fitting. The {\it Gaia} DR2 provides photometry in three bands (G, G$_{BP}$ and G$_{RP}$) plus parallax, which enables one to construct colour-magnitude diagrams (CMDs) after correcting for extinction \citep{2018A&A...616A..10G}. The use of OCs allows us to estimate the initial mass of the C~star, as long as it is a member of the cluster.

In $\S$\ref{sec2} we discuss data sources. Section \ref{sec3} describes the identification of C~stars in clusters, assesses their membership, and derives a luminosity-normalized specific frequency. In $\S$\ref{sec4} we discuss the results obtained and $\S$\ref{sec5} compares with previous work and discusses implications. 

\section{Carbon stars, clusters, and strings}
\label{sec2}
For C~stars, we considered all targets in the 3rd edition of the catalog of galactic carbon stars  \citep{2001BaltA..10....1A}, updated from \cite{1989PW&SO...3...53S}. We desired to associate as many carbon stars as possible with stellar systems of known age, namely star clusters and moving groups. Ages are typically obtained by fitting the main sequence of a color-magnitude diagram (CMD) with stellar evolutionary isochrones \citep{1995ApJ...444L...9C,Cummings_2018}. There are a number of recent papers that have used the {\it Gaia} DR2 data to identify open clusters \citep{2018yCat..36180093C, 2020A&A...635A..45C, 2019A&A...627A..35C} and also dissipated clusters \citep{2019AJ....158..122K}, where the authors employ the term "strings" to refer to the tidally smeared remnants of what once were star clusters. 

\begin{table*}
  \caption{\normalsize The means and standard deviations of position and motion parameters for the 12 clusters which have a candidate C~star member, accompanied by the values for corresponding C~stars. The clusters are arranged from youngest to oldest. \\ $^1$ $\pi$ (parallax) in mas. $^2$ $\mu_\alpha$ (proper motion in the right ascension coordinate) in mas/year. $^3$ $\mu_\delta$ (proper motion in the declination coordinate) in mas/year. $^4$ $\delta$ (declination) in degrees. $^5$ $\alpha$ (right ascension) in degrees.}
  \label{table1_stats}
    \centering
    \begin{tabular}{cm{2em}m{2em}m{2em}m{2em}m{2.25em}m{2.25em}m{2.25em}m{2.25em}m{2.25em}m{2.25em}m{2.25em}m{2.25em}m{2.25em}m{2.25em}m{2.25em}m{2.25em}}
    \hline
    Cluster & Mean $(\pi )^1$ & Std ($\pi$) & C~star ($\pi$) & Mean $(\mu_{\alpha} )^2$ & Std ($\mu_{\alpha}$) & C~star ($\mu_{\alpha}$) & Mean $(\mu_{\delta} )^3$ & Std ($\mu_{\delta}$) & C~star ($\mu_{\delta}$) & Mean $(\delta )^4$ & Std ($\delta$) & C~star ($\delta$) & Mean $(\alpha )^5$ & Std ($\alpha$) & C~star ($\alpha$) \\
    \hline
    Gulliver~29 & 0.9 & 0.06 & 0.79 & 1.31 & 0.17 & 0.84 & 2.41 & 0.11 & 2.18 & $-35.7$ & 0.1 & $-35.9$ & 256.7 & 0.47 & 258.42\\
    NGC~663 & 0.32 & 0.04 & 0.4 & $-1.11$ & 0.08 & $-1.12$ & $-0.22$ & 0.09 & $-0.49$ & 61.21 & 0.08 & 60.83 & 26.59 & 0.2 & 26.15 \\
    King~4 & 0.38 & 0.05 & 0.23 & $-0.57$ & 0.09 & $-0.73$ & $-0.07$ & 0.13 & $-0.85$ & 59.02 & 0.04 & 59.01 & 39.04 & 0.07 & 39.11\\
    Haffner~14 & 0.23 & 0.04 & 0.26 & $-1.83$ & 0.08 & $-1.69$ & 1.71 & 0.11 & 1.47 & $-28.38$ & 0.05 & $-28.4$ & 116.18 & 0.04 & 116.19\\
    Berkeley~72 & 0.13 & 0.1 & 0.21 & 0.82 & 0.19 & 0.33 & $-0.21$ & 0.13 & 0.06 & 22.25 & 0.03 & 22.27 & 87.55 & 0.02 & 87.5\\
    Berkeley~53 & 0.22 & 0.13 & 0.31 & $-3.8$ & 0.26 & $-3.87$ & $-5.68$ & 0.23 & $-6.11$ & 51.07 & 0.06 & 51.07 & 313.98 & 0.07 & 314.03\\
    NGC~2660 & 0.31 & 0.06 & 0.47 & $-2.7$ & 0.12 & $-2.69$ & 5.15 & 0.13 & 4.81 & $-47.2$ & 0.02 & $-47.21$ & 130.67 & 0.03 & 130.64\\
    FSR~0172 & 0.32 & 0.06 & 0.19 & $-2.53$ & 0.08 & $-2.444$ & $-5.93$ & 0.11 & $-5.92$ & 29.23 & 0.02 & 29.26 & 300 & 0.02 & 300.06\\
    Berkeley~9 & 0.51 & 0.08 & 0.65 & 1.52 & 0.18 & 1.74 & 0.01 & 0.13 & $-0.05$ & 52.65 & 0.03 & 52.74 & 53.17 & 0.05 & 53.23\\
    Ruprecht~37 & 0.21 & 0.1 & 0.21 & $-1.7$ & 0.14 & $-2.0$ & 2.41 & 0.11 & 2.18 & $-17.25$ & 0.01 & $-17.25$ & 117.45 & 0.02 & 117.44\\
    Berkeley~15 & 0.34 & 0.08 & 0.12 & 0.77 & 0.16 & 0.18 & $-0.87$ & 0.13 & $-1.36$ & 44.5 & 0.04 & 44.33 & 75.5 & 0.05 & 75.66\\
    Trumpler~5 & 0.28 & 0.09 & 0.41 & $-0.59$ & 0.2 & $-0.32$ & 0.28 & 0.19 & 0.51 & 9.47 & 0.1 & 9.43 & 99.13 & 0.1 & 99.14\\
    \hline
    \end{tabular}
\end{table*}

We cross-matched the \cite{2018yCat..36180093C} cluster members and the string members identified by \cite{2019AJ....158..122K} with the carbon star coordinates given in \cite{2001BaltA..10....1A} (with a search radius of 5 arcseconds) to generate a list of potential carbon star members. We then used the {\it Gaia} data itself to further assess membership via position, parallax, and proper motion. {\it Gaia} DR2 astrometry has a median uncertainty of $\approx$0.1 mas in parallax and position and $\approx$0.2 mas yr$^{-1}$ in proper motion at $G=17$ \citep{2018A&A...616A...2L}. \cite{2018yCat..36180093C} do not give cluster ages, so we adopted ages from  \cite{2013A&A...558A..53K}. For one cluster potentially containing a C~star (Gulliver~29) \cite{2013A&A...558A..53K} do not list the age or the extinction, and we adopted the age from \cite{10.1093/mnras/stz1455}. For strings, we used the ages reported in \cite{2019AJ....158..122K}.{\it Gaia} DR2 gives the line-of-sight extinction in G-band ($A_G$) and reddening ($E(BP-RP)$) for around 88 million stars with typical accuracies of order 0.46 mag in $A_G$ and 0.23 mag in $E(BP-RP)$ \citep{refId0}. We convert the $A_G$ to line-of-sight extinction in RP-band ($A_{RP}$) using
\begin{align}
\label{eq1}
    A_{RP} = A_G - 0.5*E(BP-RP)
\end{align}
We use these extinction and reddening values for individual stars while constructing the CMDs. These values are not available for all the members of the clusters or strings and thus the CMDs are produced only with stars for which these data are available from {\it Gaia} DR2. For 3 of our detected C~stars, {\it Gaia} does not provide $E(BP-RP)$ and $A_G$, so Johnson system \citep{1966CoLPL...4...99J} $E(B-V)$ given in \cite{2013A&A...558A..53K} is converted to {\it Gaia} colour excess with coefficients of \cite{10.1093/mnrasl/sly104} via
\begin{align}
\label{eq2}
    E(\zeta - \eta) = (R_\zeta - R_\eta)E(B-V)
\end{align}
where $\zeta$ and $\eta$ are any two {\it Gaia} colours and $R_\zeta$ and $R_\eta$ are their tabulated coefficients.

\section{Method}
\label{sec3}
\subsection{Matching clusters with C~stars}
\label{sec3.1}
\begin{figure}
    \centering
    \includegraphics[width=\columnwidth]{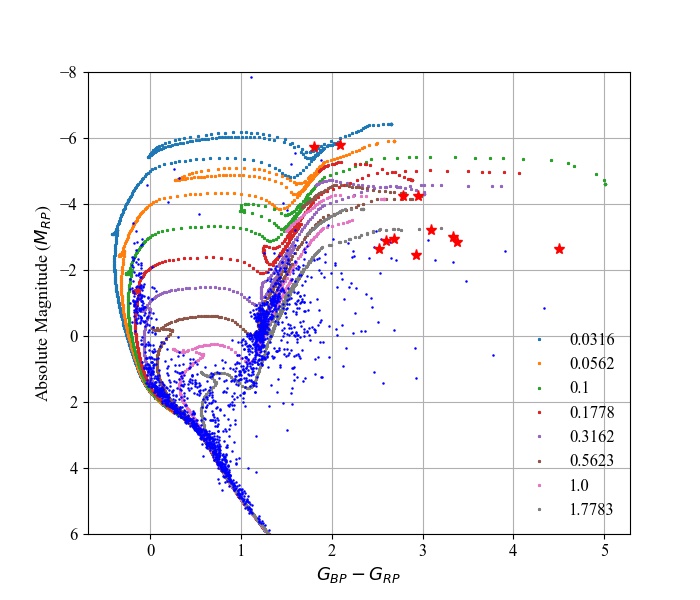}
    \caption{CMD in {\it Gaia} filters for the $12$ C~star candidates (red stars) and $4866$ other cluster members (small dots) for which extinction and reddening data are available from the {\it Gaia} DR2 catalogue. Isochrones correspond to the ages at the middle of each age bin as tabulated in Table \ref{table3_age_group} and ages are given in Gyr.}
    \label{fig1_CMD_all_C_stars}
\end{figure}

The C~star catalog \citep{2001BaltA..10....1A} was cross matched with \cite{2018yCat..36180093C} and \cite{2019AJ....158..122K} catalogs using CDS tools and a search radius of 5 arcseconds. Armed with a {\it Gaia} DR2 ID number, we found out the clusters and the strings containing the C~star candidates. As we are focusing on classical C~stars, a colour (BP-RP) range between 1.7 and 4.55 and a magnitude ($M_{RP}$) range between -2.25 and -6.0 was chosen (after correcting for extinctions in colour and magnitude). We obtained {\it Gaia} DR2 \citep{2016A&A...595A...1G} photometry, astrometry, radial velocities and derived parameters such as extinction \citep{2018A&A...616A...1G} for all the members of the clusters and stings harbouring potential C~star members. We found no classical C~star in the strings identified by \cite{2019AJ....158..122K} and obtained 12 C~stars to be potential members of clusters identified by \cite{2018yCat..36180093C}.These C~star candidates were confirmed by  literature searches, spectral type, and their placement in the CMD. All candidates appear to be bona fide C~stars and the 12 C~star candidates are shown in Fig. \ref{fig1_CMD_all_C_stars}.

\begin{figure*}
    \centering
    \includegraphics{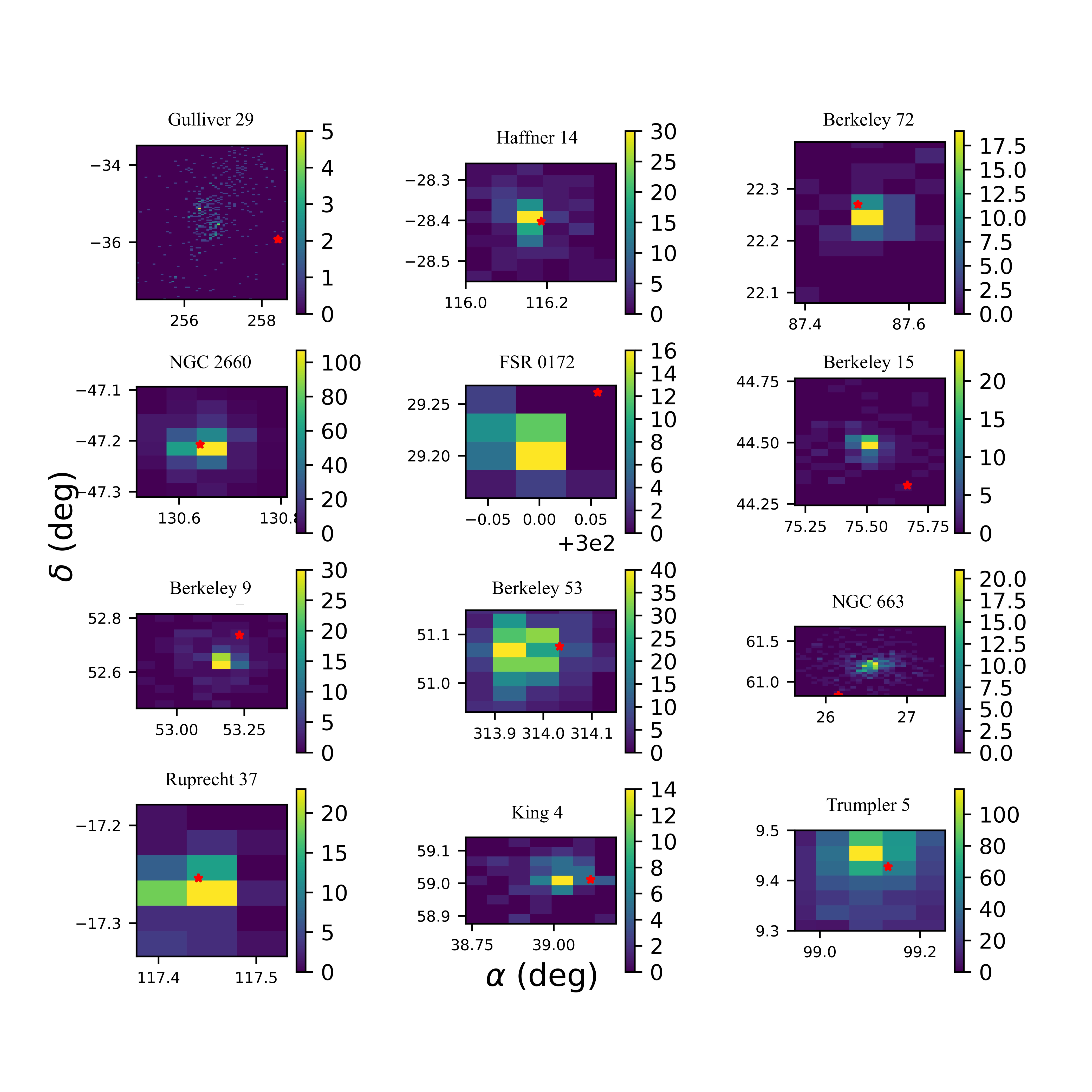}
    \caption{\normalsize The sky position of candidate C~stars (star) in respective clusters (image). The colour scale signifies Cantat-Gaudin et al. (2018) number density. }
    \label{fig4_ra_dec_all}
\end{figure*}

To assess if a C~star is a member of its cluster or string, we looked into the location of the star in proper motion ($\mu_\alpha , \mu_\delta$), position ($\alpha , \delta$), and parallax ($\pi$) spaces relative to the distributions defined by the clusters. We were liberal in our inclusion for two reasons. Firstly, \textit{Gaia} errors are larger for carbon stars due to their large angular size and photocentric variability \citep{2011A&A...528A.120C,2018A&A...617L...1C}, and secondly, the \cite{2018yCat..36180093C} technique may miss valid members \citep{mahmudunnobe2021membership}. In particular, we consulted sky survey images in addition to Figure \ref{fig4_ra_dec_all}, which plots member density on the sky.

\begin{figure*}
    \centering
    \includegraphics{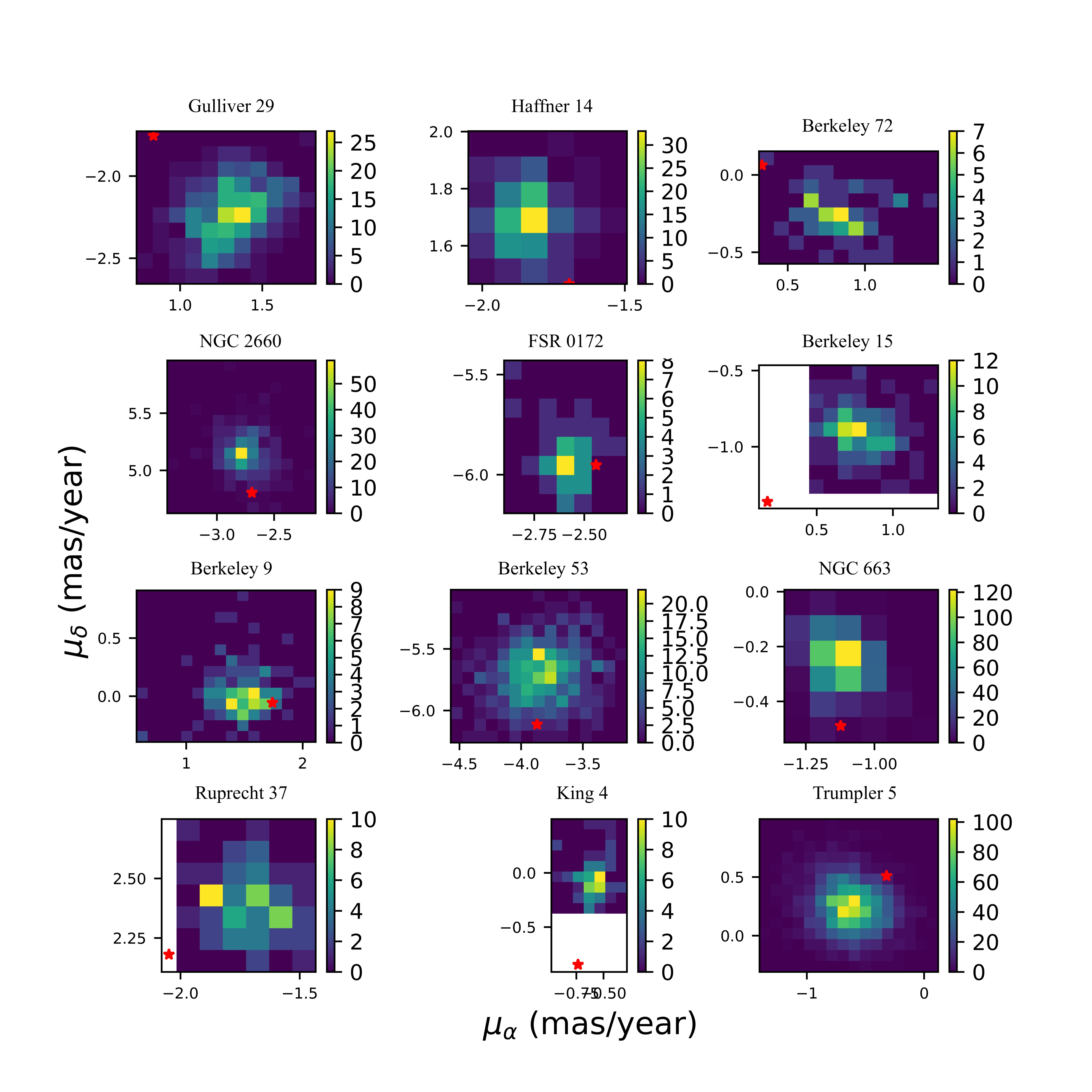}
    \caption{\normalsize The candidate C~stars (star) and other cluster stars (image) in $\mu_\alpha-\mu_\delta$ space. The colour scale tracks number density of stars.}
    \label{fig2_proper_motions_all}
\end{figure*}
\begin{figure*}
    \centering
    \includegraphics{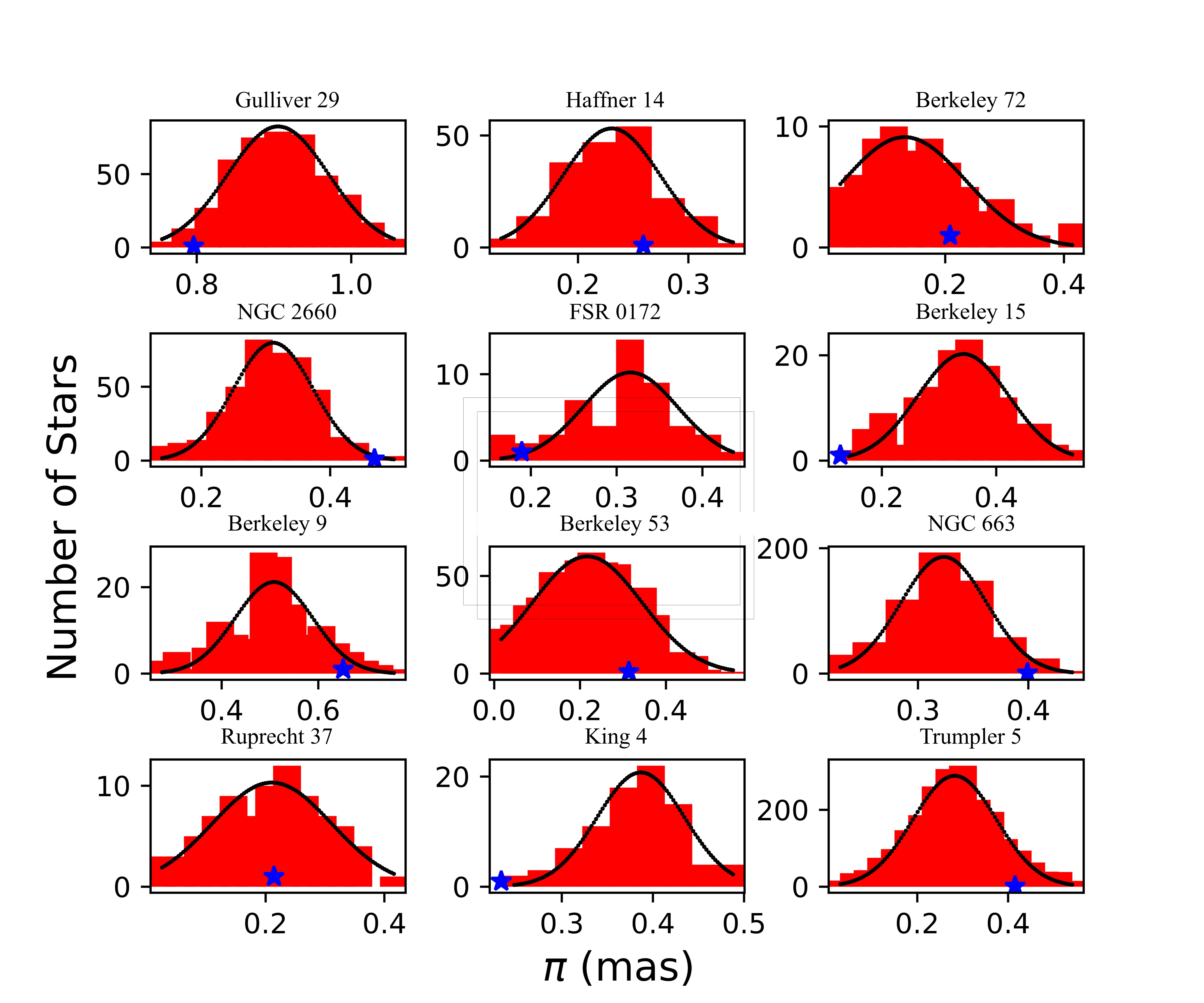}
    \caption{\normalsize The $\pi$ of candidate C~stars (star) among respective cluster stars (histogram).}
    \label{fig3_parallax_all}
\end{figure*}

Distributions in Table \ref{table1_stats} shows the means and standard deviations of distributions of all 12 clusters in position, proper motion, and distance along with the corresponding values for the C~stars possibly associated with those clusters. For Table \ref{table1_stats} Gaussian fits, we considered \cite{2018yCat..36180093C} members that have membership probability $>50\%$. 

\textbf{Excluded stars.} We rejected V617~Sco as a member of Gulliver~29 because it lay far from the cluster in sky position (Fig. \ref{fig4_ra_dec_all}), proper motion (Fig. \ref{fig2_proper_motions_all}), and distance (Fig. \ref{fig3_parallax_all}). Also, the theoretically predicted mass limit for the formation of C~stars is $~8M_\odot$ \citep{1999A&A...346..805V, 2017A&A...600A..92R}. Gulliver~29 is extremely young ($0.0158$ Gyr) which translates to an initial mass of $~11M_\odot$. The \cite{2018yCat..36180093C} paper assigns a probability of just $0.1$ for this particular star. 
Carbon star NIKC~2-30 ({\it Gaia} DR2 464249891274666752) is $\sim$4 M$_\odot$ if it is a member of King~4. We reject NIKC~2-30 as a member of the King~4 OC based on the fact that it lies far outside the $\mu_\alpha$-$\mu_\delta$ (Fig. \ref{fig2_proper_motions_all}) and distance (Fig. \ref{fig3_parallax_all}) cluster star distributions. It is also reported to have a membership probability of 0.1 by \cite{2018yCat..36180093C}. 
If NIKC~3-82 ({\it Gaia} DR2 2055750405085315) is a member of Berkeley~15, it is of normal mass for a C-star. However, it is also rejected as a member based on position, proper motion, and distance (Figs. \ref{fig4_ra_dec_all}, \ref{fig2_proper_motions_all}, \ref{fig3_parallax_all}). 

\begin{figure*}
    \centering
    \includegraphics{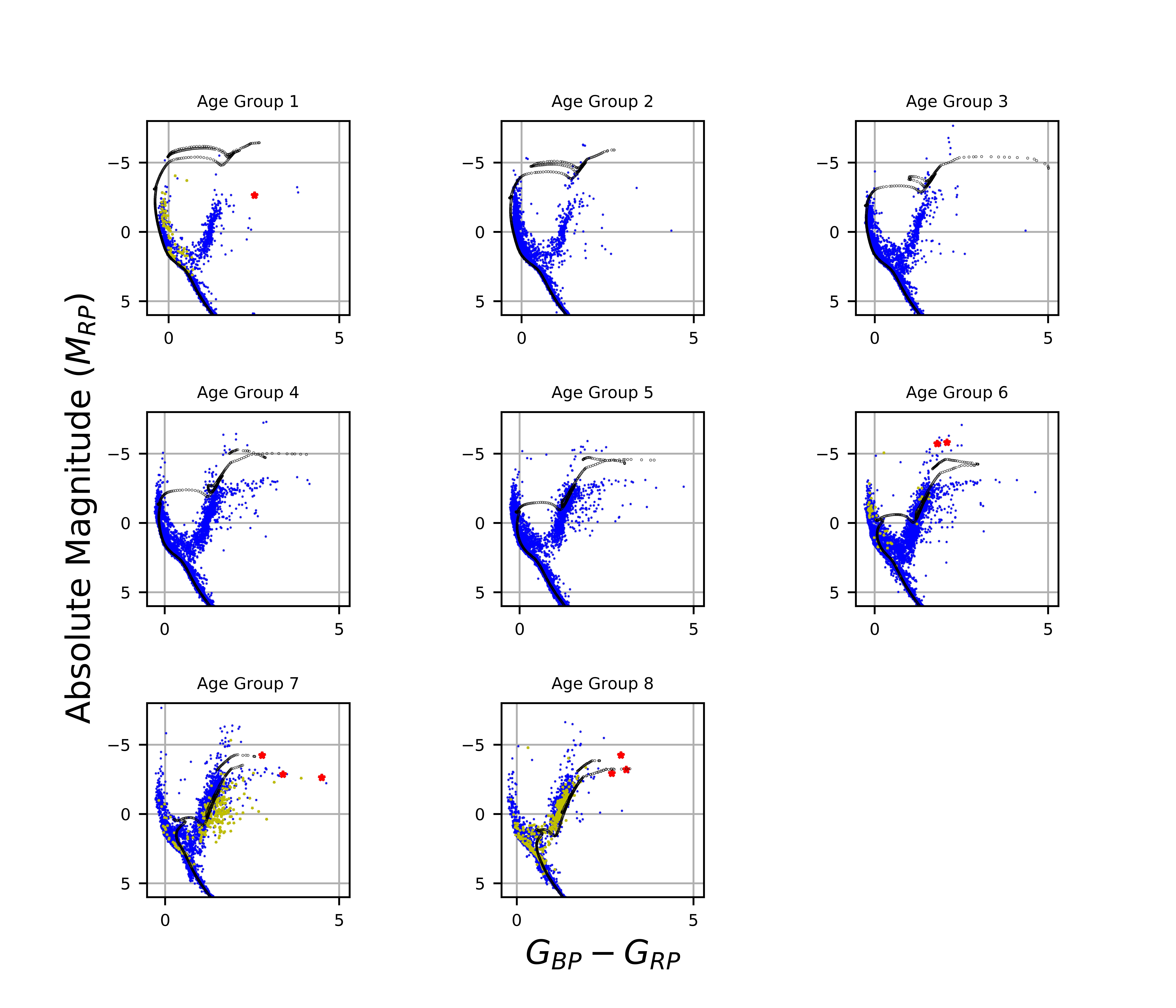}
    \caption{\normalsize CMD for the clusters belonging to different age groups. Each isochrone here refers to an age which is midpoint of each age-bin (age-bins described in Table \ref{table3_age_group}). Stars with membership probability $>$50$\%$ by Cantat-Gaudin et al. (2018) are plotted. Despite this, one can readily spot field star contamination, especially of first-ascent giant stars, though the  percentage of such contamination is low. C~stars are plotted as red asterisks and the yellow circles denote members of OCs which harbor the C-stars.} 
    \label{fig5_cmd_all}
\end{figure*}

None of the smattering of luminous red stars in the CMDs of Figs. \ref{fig1_CMD_all_C_stars} and \ref{fig5_cmd_all} that lie near the confirmed C~stars could be confirmed as C~stars. One comes close: \textit{Gaia} DR2 213087625504802304, a probable member of NGC 1798 (age group 7 as defined below), is an S~star, intermediate between types M and C.

\textbf{Questionable stars.} We include star Case~49 ({\it  Gaia} DR2 509727788151134720; USNO-B1.0 1508-0065037) as a member of NGC~663 even though it shows ambiguity. It shows a typical optical variability of around 0.2 magnitude and is regarded as an irregular variable\citep{article}. NGC 663 suffers from variable reddening and appears to have a mixture of ages present, all less than 50 Myr \citep{2005MNRAS.358.1290P}. Case 49 lies among NGC 663 cluster stars in proper motion (Fig. \ref{fig2_proper_motions_all}) and distance (Fig. \ref{fig3_parallax_all}). However, it lies well away from cluster center in sky position (Fig. \ref{fig4_ra_dec_all}) and for its high mass ($\sim$7 M$_\odot$), it is underluminous in the CMD (Fig. \ref{fig5_cmd_all}). Its low luminosity resembles similar stars in next-door NGC 654 \citep{2005MNRAS.358.1290P} but cannot easily be explained by self-extinction due to mass loss because its mass-loss (in the form of dust) rate is relatively modest $0.7\times 10^{-9}$ M$_\odot$ yr$^{-1}$ \citep{2000A&A...357..225J}. However, it does have a high total (both gas and dust) mass-loss rate of $1.1\times10^{-6}$ M$_\odot$ yr$^{-1}$ \citep{refId0}. It is possible that significant mass loss has occurred in this star, lowering its luminosity and enabling the revelation of C-rich layers. In summary, we doubt that Case~49 is a member of NGC~663, but we carry it along in our analysis anyway. Because its mass is so high, it is easily ignored in the figures that follow.

C*~908 ({\it Gaia} DR2 5718601035951388800) lies in the field of Ruprecht~37 and appears to have the same distance. Only its proper motion is toward the edge of the distribution. Ruprecht~37 is not a rich cluster, but it lies in a rich field at $b=4.5^{\circ}$. The star itself is very red and of the correct luminosity. It does not yet have a spectral type, but it has appeared in all editions of the carbon star catalog \citep{2001BaltA..10....1A}. We include it in our analysis.

\textbf{Members.} Table \ref{table1_stats} accompanied by Fig. \ref{fig4_ra_dec_all}, Fig. \ref{fig2_proper_motions_all}, and Fig. \ref{fig3_parallax_all} show that the remaining C~stars are high probability members. Fig. \ref{fig5_cmd_all} shows in age group 6 that BM~IV~34 and Case~121 appear overluminous, but this might be an overcorrection for extinction and not necessarily an age effect. Table \ref{table2_C_stars} gives the {\it Gaia} IDs and common names along with the respective cluster names and ages for the 9 C~stars that are being analysed in this paper (including the borderline cases of Case~49 and C*~908). 

\begin{table}
  \caption{Details of the 9 C~stars possibly connected with OCs. Common names obtained from SIMBAD database. Metallicity of the clusters are also listed.}
  \label{table2_C_stars}
    \centering
    \begin{tabular}{m{8.5em}m{3.75em}m{3.75em}m{2.75em}m{1.75em}m{1.5em}}
    \hline
    {\it Gaia} ID & Common Name & Cluster & [Fe/H] & log[age (yr)] & E(B-V) (mag)\\
    \hline
    509727788151134720 & Case~49 & NGC~663 & -0.7\tablefootnote[1]{\cite{2010A&A...517A..32P}} & 7.5 & 0.94 \\
    5599918758725999616 & BM~IV~34 & Haffner~14 & ... & 8.79 & 0.64 \\ 
    3424247755746068224 & Case~121 & Berkeley~72 & -0.24\tablefootnote[2]{\cite{2008PASJ...60.1267H}} & 8.83 & 0.90\\
    2169782297869982464 & Case~473 & Berkeley~53 & 0.00\tablefootnote[3]{\cite{Donor_2018}} & 9.1 & 2.03 \\
    5329370041381407232 & BM~4~90 & NGC~2660 & 0.04\tablefootnote[4]{\cite{2013A&A...558A..53K}} & 9.1 & 0.47\\
    2030043269178382848 & IRAS~19582 +2907 & FSR~0172 & ... & 9.12 & 1.09 \\
    443239667174584704 & BI~Per & Berkeley~9 & -0.17\tablefootnote[5]{\cite{2019A&A...623A..80C}} & 9.3 & 1.0 \\
    5718601035951388800 & C*~908 & Ruprecht~37 & -0.3\tablefootnote[6]{\cite{article}} & 9.3 & 0.0\\
    3326781272625443712 & V493~Mon & Trumpler~5 & -0.3\tablefootnote[4]{\cite{2013A&A...558A..53K}} & 9.5 & 0.84 \\
    \hline
    \end{tabular}
\end{table}

\subsection{Number to luminosity ratio}
\label{sec3.2}

The specific frequency of C~stars, the number normalized by cluster luminosity, was compiled for the Magellanic Cloud clusters by \cite{2007A&A...462..237G}. This is a useful quantity for population studies because it gives relatively direct information about stellar evolutionary lifetimes for stars in the carbon-dominant stages near the ends of their lives. While some star clusters in the Magellanic Clouds contain many C~stars, our search for Galactic C~stars yielded at most one C~star per cluster. It is important for proper accounting to include cluster light from the clusters that yielded zero C~stars. Under the reasonable assumption that the C~star catalog was approximately magnitude-limited, we limited the cluster sample to lie within the distance of the farthest confirmed C~star member. The most distant C~star is in cluster FSR~0172 with $\pi=0.19$ mas which translates to a distance of $5.2$ kpc with an uncertainty of $2.7$ kpc (considering a median uncertainty of $\approx 0.1$ mas in parallax values for {\it Gaia} DR2 \citep{2018A&A...616A...2L}) . 

Throughout, we employ Padova isochrones \citep{2008A&A...482..883M} updated through 2011 with a \cite{2001MNRAS.322..231K} initial mass function incorporated via the \cite{1994ApJS...95..107W} models. Low mass evolution is added from \cite{2011ApJS..192....3P} and the low-mass cutoff is 0.08 M$_\odot$. {\it Gaia} photometry is synthetic using passbands from \cite{2018A&A...616A...4E} and assumed Vega colours of zero. We also used the same Vega spectrum as \cite{2018A&A...616A...4E} to assign photometric zeropoints for {\it Gaia} blue ($G_{BP}$) and red ($G_{RP}$) filters. For the Johnson $V$ filter, we use the observed colours and bolometric corrections of Vega \citep{1964BOTT....3..305J}. 
 
We arrange age bins evenly spaced in log[age (yr)] from 7.5 to 9.5 with an increment of 0.25. For each bin, the models yield a main sequence turnoff mass, number counts, integrated luminosities, and mass-to-light ratios in all relevant passbands. For each cluster, we distance-correct and account for dust extinction before comparing with models. To estimate total luminosity, we count main sequence stars with $G_{BP} - G_{RP} <$ 0.9 mag and more luminous than the magnitude threshold listed in Table \ref{table3_age_group}. We treat the theoretical Hess \citep{1923AN....220...65H} diagram the same way, then scale to find each cluster's mass, luminosity, or star count under the assumption of a Kroupa IMF.

Table \ref{table3_age_group} lists our binning scheme, a main sequence turnoff mass corresponding to the midpoint of the bin, colour cutoffs and $M_{RP}$ cutoffs for number counting. Note that for age groups 2, 3, 4 and 5 no C~star was detected, so no analysis is required for those two age bins. 

\begin{table}
  \caption{Parameters for age bins. The number of OCs in each age bin is obtained from Cantat-Gaudin et al. (2018). } 
  \label{table3_age_group}
    \centering
    \begin{tabular}{@{}cm{2.5em}cm{2.75em}m{6em}m{2.75em}}
    \hline
    Age group & Number of OC & log(age) & $M_{TO}$ (M$_\odot$) & Colour ($G_{BP}-G_{RP}$) cutoff & $M_{RP}$ cutoff \\
    \hline
    1 & 60 & 7.50-7.75 & 6.79 & 0.9 & 2 \\    
    2 & 86 & 7.75-8.00 & 5.20 & N/A & N/A \\
    3 & 96 & 8.00-8.25 & 4.10 & N/A & N/A \\
    4 & 123 & 8.25-8.50 & 3.23 & N/A & N/A \\
    5 & 183 & 8.50-8.75 & 2.57 & N/A & N/A \\
    6 & 193 & 8.75-9.00 & 2.09 & 0.9 & 4 \\
    7 & 130 & 9.00-9.25 & 1.67 & 0.9 & 4 \\
    8 & 58 & 9.25-9.50 & 1.38 & 0.9 & 4 \\
    \hline
    \end{tabular}
\end{table}

We approximate a volume-limited sample by omitting clusters farther than the farthest C~star in our list, 5.2 kpc. The C~stars are identified from the spectrum of the particular star \citep{2004AJ....127.2838D, 2002AJ....124.1651M}, usually down to a magnitude limit. The probability of detecting and confirming C~stars using spectroscopic methods decreases with distance and does not span the Galaxy. It is thus best to leave out clusters beyond the farthest C~star as some might very well harbour undetected C~stars. Within the distance limit, we account for all the luminosity of clusters stars, whether they contain a C~star or not. Once armed with the summed luminosity of all clusters in each age bin, it is straightforward to take the ratio of the number of carbon stars to the summed luminosity. We have assigned a zero value to this ratio in each of the bins where we do not have any C~star.

\section{Results}
\label{sec4}

As outlined in Sec. \ref{sec3.2}, we divide the ages into logarithmic bins of width 0.25 spanning 7.5 $\leq$ log(age) $\leq$ 9.5. Integrated luminosities and isochrone number counts for the Worthey model default population mass of $10^6$ M$_\odot$ were calculated at age bin centers. For Johnson/Cousins V and {\it Gaia} RP-band we assumed solar absolute magnitudes $M_{V,\odot}$ = 4.84 and $M_{RP,\odot}$ = 4.229. A scaling factor is calculated by dividing the total number of model stars in the population by the number of stars brighter than some cutoff magnitude (listed in Table \ref{table3_age_group}). For all clusters in each age bin, a count of observed cluster members
above the magnitude cutoff was made and then these counts were summed
and multiplicatively scaled to total luminosities L$_V$ and L$_{RP}$ for the
combined set of clusters in the age bin. We then compute the specific frequency of C-stars (number per unit luminosity) in each bin. Table \ref{table4_lum_ratio} summarises the results of this exercise as a function of main sequence turn-off mass, $M_{TO}$ (M$_\odot$).

\begin{table}
  \caption{$N_C/L_V$ and $N_C/L_{RP}$ ratio tabulated as a function of $M_{TO}$ ($M_\odot$).}
  \label{table4_lum_ratio}
    \centering
    \begin{tabular}{@{}llllcc}
    \hline
    $M_{TO}$    & $N_C$ & $L_V$             & $L_{RP}$            & $N_C/L_V$                   & $N_C/L_{RP}$ \\
    ($M_\odot$) &       & ($10^6$ L$_\odot$) & ($10^6$ L$_\odot$) & $(10^{-6}$ L$_\odot^{-1}$ ) &  $( 10^{-6}$ L$_\odot^{-1} )$ \\
    \hline
    6.79 & 1 & 0.31 & 0.07 & 3.23 & 14.28 \\
    5.2 & 0 & 1.28 &  0.26 & 0 & 0 \\
    4.1 & 0 & 0.88 & 0.17 & 0 & 0 \\
    3.23 & 0 & 1.70 & 0.33 & 0 & 0 \\
    2.57 & 0 & 1.41 & 0.25 & 0 & 0 \\
    2.09 & 2 & 0.27 & 0.04 & 7.41 & 50 \\
    1.67 & 3 & 0.15 & 0.015 & 20 & 200 \\
    1.38 & 3 & 0.05 & 0.004 & 60 & 750 \\
    \hline
    \end{tabular}
\end{table}

\begin{figure}
    \centering
    \includegraphics[width=\columnwidth]{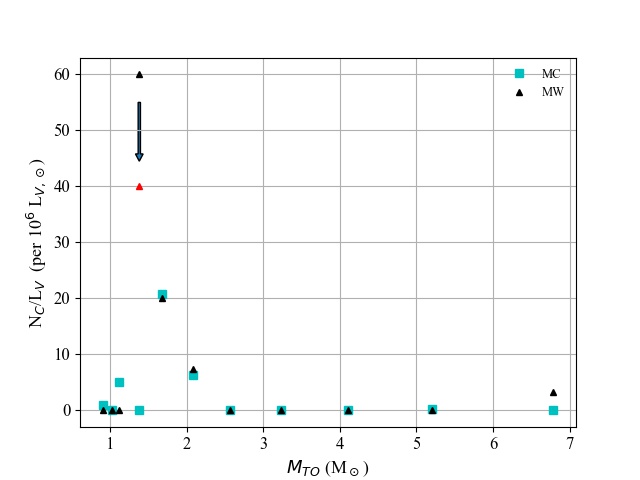}
    \caption{Number of C~stars normalized by population luminosity in the $V$ band as a function of main sequence turnoff mass. Milky Way (triangles) and Magellanic Cloud (squares) C~stars become more common below 2.5 M$_\odot$. The red triangle shows the ratio if C*~908 is not considered a member of Ruprecht~37 and the arrow shows the shift in the ratio.}
    \label{fig6}
\end{figure}

\begin{figure}
    \centering
    \includegraphics[width=\columnwidth]{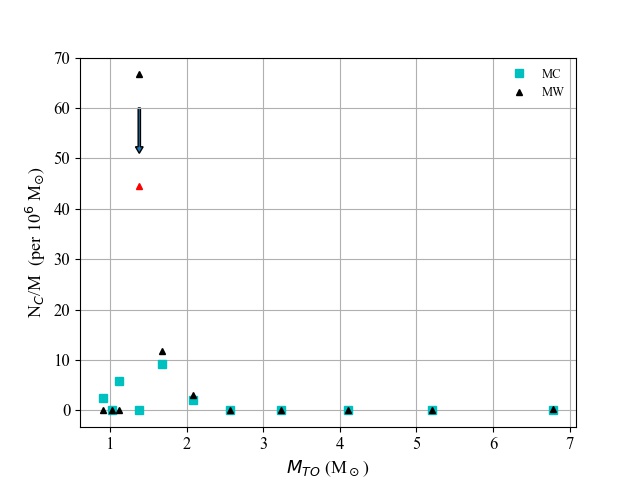}
    \caption{Milky Way (triangles) and Magellanic Cloud (squares) carbon star numbers normalized by population mass as a function of main sequence turnoff mass. The red triangle shows $N_C/M$ if C*~908 is not considered a member of Ruprecht~37. Masses were derived from Fig. \ref{fig6} via theoretical abundance-sensitive mass to light ratios. We assumed a Kroupa IMF, that local stars have [M/H] = 0, and that MC clusters have $<$[M/H]$>$ $= -0.53$.}
    \label{fig7}
\end{figure}

\begin{figure}
    \centering
    \includegraphics[width=\columnwidth]{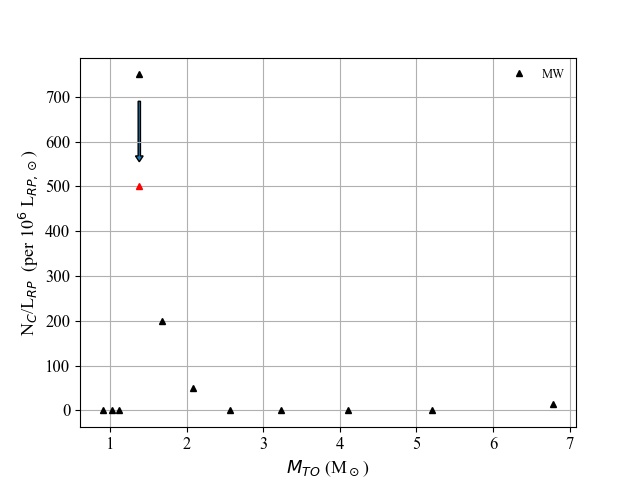}
    \caption{Number of carbon stars normalized by population luminosity in the {\it Gaia} RP band as a function of main sequence turnoff mass. The red triangle is for the case if C*~908 is not considered a member of Ruprecht~37.}
    \label{fig8}
\end{figure}

Fig. \ref{fig6} shows the ratio of number of C~stars to the luminosity in V-band for both the Milky Way (MW) (this work) and the Magellanic Cloud (MC) \citep{2007A&A...462..237G}. In Fig. \ref{fig7}, this number to luminosity (in V-band) ratio ($N_C/L_V$) is converted to number to mass ($N_C/M$) ratio using a theoretically predicted mass to luminosity (in V-band) ratio ($M/L_V$). It is to be noted that all the calculations and comparisons are based on OC data only. Fig. \ref{fig8} shows the number to luminosity (in {\it Gaia} RP-band) ratio in the {\it Gaia} RP-band. We find exactly the same trend as in Fig. \ref{fig6} but with different values as it is using a different passband.

\section{Discussion and conclusion}
\label{sec5}

In this study, we found 12 C~stars potentially associated with 12 different open clusters in the Milky Way after cross referencing the catalogues of open clusters by \cite{2018yCat..36180093C} and by \cite{2013A&A...558A..53K} with the C~star catalogue by \cite{2001BaltA..10....1A}. We chose members that have been assigned a probability of greater than 50\% by \cite{2018yCat..36180093C} although \cite{mahmudunnobe2021membership} points out that \cite{2018yCat..36180093C} might be under-reporting the members of the clusters. We discarded $\sim 14$ M$_\odot$ V617~Sco along with NIKC~2-30 and NIKC~3-82 as cluster members based on position, proper motion and parallax data. We retained $\sim 7$ M$_{\odot}$ Case~49 although it is a borderline member of NGC~663. We retained C*~908 as a member of Ruprecht~37. Although one component of its proper motion put it at the fringe of the cluster distribution, its other data was fully consistent with membership. With a culled list of 9 C~stars of known initial mass we calculated the number to luminosity ratio in V-band and {\it Gaia} RP-band as a function of main sequence turn-off mass and equivalent cluster age. Above a turn-on mass of 1.24 M$_\odot$, we see a trend of decreasing C~star fraction with $M_{TO}$ to something near zero for $M_{TO} > 2.5 M_\odot$. 

We transformed the $N_C/L_V$ ratio for the MCs \citep{2007A&A...462..237G} to our binning scheme by tracking numbers of C~stars and luminosities cluster by cluster and rebinning. In comparing our results with those of \cite{2007A&A...462..237G} (Fig. \ref{fig7}), we see only one point of dissimilarity. In the M$_{TO} = 1.38M_\odot$ bin there are no MC C~stars, but quite a high fraction in the Milky Way. This is based on three stars, so it is likely to be real. 

As for high-mass C~stars, we note that \cite{2007A&A...462..237G} reports a high mass C~star at $\sim6$~M$_\odot$. Our highest mass is $\sim7$~M$_\odot$ for questionable member Case~49. Whether or not the mass of this star is secure, it is seen in Fig. \ref{fig7} that its impact is minimal on the near-zero number fraction. In terms of stellar evolution theory, the upper limit for C~star production is of interest because models must be made that can dredge up carbon to the surface even at high stellar masses.

We assume that the driving difference between the solar neighborhood and the Magellanic Clouds is heavy element abundance. \cite{2007A&A...462..237G} in their Table 2 list metallicities for each of their age bins. We take the average of those values to get the metallicities of LMC and SMC to be [Fe/H] $= -0.33$ dex and [Fe/H] $= -0.93$ dex, respectively (note that we leave out the metallicity of their last age bin for LMC as it is reported to be $<-1.5$ dex). At low metallicity, there are fewer O atoms, and therefore it takes fewer dredged-up C atoms to attain number dominance at the surface. Metal-poor populations may therefore find it easier to produce C~stars and there might be more of them. Fig. \ref{fig7} presents slim evidence for a metallicity effect \textit{except} in the M$_{TO} = 1.38 M_\odot$ bin, where the opposite is evident (at face value; see below). In that single bin, lower-mass solar metallicity stellar populations appear to produce more C~stars.

What about other low-mass, solar-metallicity sites? M31 satellite galaxies NGC~205 and M~32 may have young stars, but the bulk of their stellar populations are older than log(age) = 9.5  \citep{2004AJ....128.2826W, 2004PASP..116..295W} and M~32, at least, is closer to solar metallicity than the MCs \citep{1996AJ....112.1975G, 2011ApJ...727...55M}, though NGC~205 may be comparable \citep{2014MNRAS.444.1705G}.  \cite{2016ApJ...828...15H} identified C~stars in M31 satellites using criteria from \cite{2015ApJ...810...60H}. They report 7 and 5 C~stars for NGC~205 and M~32, respectively. Using photometric and distance data from the NED database we find that $N_C/L_V = 2\times10^{-8}$ $L_{V,\odot}^{-1}$ and $0.59\times10^{-8}$ $L_{V,\odot}^{-1}$  for NGC~205 and M~32, respectively. These near-zero values imply that we should expect near-zero C~star production for stars of 1 or 1.1 M$_\odot$. 

We also calculate the number to luminosity ratio at a distance of 10 kpc from the center of M~31, roughly analogous to the solar circle although the metallicity at this location appears to be slightly lower; [Fe/H]=$-$0.5 dex \citep{2014ApJ...796...76G}, but which is similar to that of MW at 10 kpc from the center \citep{2010AAS...21531001C}. We use bricks 15 \& 16 as described in \cite{2019ApJ...879..109B} which corresponds to brick 15 in \cite{2012ApJS..200...18D}. The angular distance between center of brick 15 \citep{2012ApJS..200...18D} and that of the M~31 nucleus is $\approx 50'$. Fig. 12 of \cite{1987A&AS...69..311W} gives a surface brightness. Using the angular area of two bricks in \cite{2019ApJ...879..109B} (each of size $136'' \times 123''$) and a NED distance we find a $V$ luminosity with which we can normalize the C~star counts from \cite{2019ApJ...879..109B}. The 59 C~stars reported normalize to $N_C/L_V = 2\times 10^{-6}$  $L_{V,\odot}^{-1}$.

While we do not know the distribution of turnoff masses at the location we sampled in M31, the result of $N_C/L_V = 2\times 10^{-6}$  $L_{V,\odot}^{-1}$ is in good enough accord with MW and MC C~star frequencies (Fig. \ref{fig6}) to raise no concerns about unexplained C~star frequencies. 

We conclude that at intermediate masses and within counting statistics there is no detected metallicity effect for C~star production over the factor-of-several metallicity difference between the MCs and the MW. Neither increased mass-loss for metal-rich populations nor lengthened timescales for metal-rich populations appear to affect C~star production (or else the two effects compensate for each other). In our lowest-mass bin ($\approx 1.38$ M$_\odot$), however, there is a striking difference in that these stars have a strong C~star phase at solar metallicity but none at all at MC metallicities. 

Rather than posit some astrophysical effect, however, one must remember that Fig. 1 of \cite{1996A&A...316L...1M} shows no calibrating clusters between turnoff masses of 1.25 and 1.7 M$_\odot$, corresponding to an age range of roughly 1.6 to 4 Gyr where no data exists. A lone cluster at age $\sim$4 Gyr shows only M stars and no C stars. In fact, our mass bin at 1.38 M$_\odot$ is simply unrepresented in the MC, and our humble three-star sample is the first empirical data to fall into this age range.

V493 Mon in Trumpler~5 deserves a note. This cluster is probably metal poor by a factor of two compared to solar \citep{2003JKAS...36...13K,2004MNRAS.349..641P}, so it may not address our goal of studying metal-rich populations perfectly. Furthermore, while we assume an age of 3 Gyr, other estimates vary. \cite{2003JKAS...36...13K} quotes 2.4 Gyr, while \cite{2004MNRAS.349..641P} put it at 5.0 Gyr. If the latter estimate holds true, it pushes the lowest mass that can produce a carbon star lower. Modelers generally cut off carbon star production for ages older than 3 Gyr (e.g., \cite{2005MNRAS.362..799M}), so if production lingers later, it has implications for integrated light studies.

Extrapolating to the next-lower bin in mass (9.50 $<$ log age $<$ 9.75; $\sim 4$ Gyr; $M_{TO} = 1.24 M_\odot$), C~star production drops to zero, as exemplified by our findings, the results from NGC~205 and M~32, and the MC clusters. None of the old open clusters in the MW such as M~67, NGC~188, or NGC~6791 contain classical C~stars.

\section{Data availability}

The data underlying this article are generally publicly available. This work has made use of data from the European Space Agency (ESA) mission {\it Gaia} (\url{https://www.cosmos.esa.int/gaia}), processed by the {\it Gaia}
Data Processing and Analysis Consortium (DPAC, \url{https://www.cosmos.esa.int/web/gaia/dpac/consortium}). Funding for the DPAC has been provided by national institutions, in particular the institutions participating in the {\it Gaia} Multilateral Agreement. This research also made use of the cross-match service provided by CDS, Strasbourg and the SIMBAD database, operated at CDS, Strasbourg, France.  This research has made use of the NASA/IPAC Extragalactic Database (NED), which is funded by the National Aeronautics and Space Administration and operated by the California Institute of Technology. Any additional data underlying this article will be shared on reasonable request to the corresponding author.

\bibliographystyle{mnras}
\bibliography{cstar} 

\bsp	
\label{lastpage}

\end{document}